\documentclass[twocolumn]{aastex631}
\usepackage{CJK}

\shorttitle{Recognizing Blazars Using Radio Morphology from the VLA Sky Survey}
\shortauthors{Xie et al.}

\begin{document}

\title{Recognizing Blazars Using Radio Morphology from the VLA Sky Survey}

\begin{CJK*}{UTF8}{gbsn}
\author[0000-0002-0125-6679]{Zhang-Liang Xie (谢彰亮)}
\affiliation{Max Planck Institut für Astronomie, Königstuhl 17, D-69117, Heidelberg, Germany}

\author[0000-0002-2931-7824]{Eduardo Ba\~{n}ados}
\affiliation{Max Planck Institut für Astronomie, Königstuhl 17, D-69117, Heidelberg, Germany}

\author[0000-0003-4747-4484]{Silvia Belladitta}
\affiliation{Max Planck Institut für Astronomie, Königstuhl 17, D-69117, Heidelberg, Germany}
\affiliation{INAF, Osservatorio di Astrofisica e Scienza dello Spazio, Via Gobetti 93/3, 40129 Bologna, Italy}

\author[0000-0002-5941-5214]{Chiara Mazzucchelli}
\affiliation{Instituto de Estudios Astrofísicos, Facultad de Ingeniería y Ciencias, Universidad Diego Portales, Avenida Ejercito Libertador 441, Santiago, Chile}

\author[0000-0002-4544-8242]{Jan-Torge Schindler}
\affiliation{Hamburger Sternwarte, Universität Hamburg, Gojenbergsweg 112, D-21029 Hamburg, Germany}

\author[0000-0003-0821-3644]{Frederick Davies}
\affiliation{Max Planck Institut für Astronomie, Königstuhl 17, D-69117, Heidelberg, Germany}

\author[0000-0001-9024-8322]{Bram P. Venemans}
\affiliation{Leiden Observatory, Leiden University, PO Box 9513, NL-2300 RA Leiden, The Netherlands}

\begin{abstract}
Blazars are radio-loud Active Galactic Nuclei (AGN) whose jets have a very small angle to our line of sight. Observationally, the radio emission are mostly compact or a compact-core with a 1-sided jet. With 2.5$^{\prime\prime}$ resolution at 3 GHz, the Very Large Array Sky Survey (VLASS) enables us to resolve the structure of some blazar candidates in the sky north of Decl.\ $-40$ deg. We introduce an algorithm to classify radio sources as either blazar-like or non-blazar-like based on their morphology in the VLASS images. We apply our algorithm to three existing catalogs, including one of known blazars (Roma-BzCAT) and two of blazar candidates identified by WISE colors and radio emission (WIBRaLS, KDEBLLACS). We show that in all three catalogs, there are objects with morphology inconsistent with being blazars. Considering all the catalogs, more than 12\% of the candidates are unlikely to be blazars, based on this analysis. Notably, we show that 3\% of the Roma-BzCAT ``confirmed” blazars could be a misclassification based on their VLASS morphology. The resulting table with all sources and their radio morphological classification is available online.

\end{abstract}

\section{Introduction}

 Blazars are a subset of AGN that have radio jets pointed towards us at a very small angle ($\theta<20^{\circ}$, \citealt{Kollgaard1992, Urry1995}). Despite being one of the rarest sub-classes of AGN, they are the most commonly found sources  in the $\gamma$-ray sky (e.g., \citealt{Hartman1999, Ajello2020}). Blazars are highly variable and polarized radio sources that exhibit compact morphology. Additionally, they are luminous across a wide range of frequencies, spanning from radio to $\gamma$-ray \citep{B&A1979, Abdo2010}. Observational signs of jets can be found across a broad range of the electromagnetic spectrum \citep{Blandford2019}. In particular, blazars can be classified into two sub-classes based on their optical/near-infrared emission lines: BL Lacertae objects (BL Lac), which have either no or very weak emission lines (with rest-frame equivalent width $<$5\,\AA; \citealt{Stickel1991}), and flat spectrum radio quasars (FSRQ), which have broad optical/near-infrared emission lines similar to those of Type-1 quasars and a flat radio spectrum \citep{Sambruna1996}.
 
 A large sample of blazars is necessary to study their radiation mechanism and relativistic jet beaming effects. However, confirming a large number of blazars is time and resource-consuming (e.g., \citealt{Shaw2012}). The Roma-BzCAT is currently the most comprehensive collection of confirmed blazars, containing $3,561$ sources that are classified as BL Lac or FSRQ \citep{Massaro2015}. All blazars in Roma-BzCAT must meet the criteria of having spectroscopic information showing characteristic features of their class, at least one radio detection exhibiting compact morphology or one-sided jet and an isotropic X-ray luminosity close to or higher than $10^{43}\, \mathrm{erg}\, \mathrm{s}^{-1}$ \citep{Massaro2009}.

In an effort to expand the pool of identified blazars while minimizing the presence of potential contaminants, primarily other forms of AGN, \cite{D'abrusco2012} studied the mid-infrared (MIR) colors of established blazars, employing the Wide-Field Infrared Survey Explorer (WISE, \citealt{Wright2010}) to uncover a distinctive color region in which blazars reside. \cite{D'Abrusco2014} assembled a catalog of blazar candidates (WIBRaLS) based on their WISE IR colors (W1-W2, W2-W3, W3-W4), while requiring the existence of a radio counterpart either in the NRAO VLA Sky Survey (NVSS, \citealt{Condon1998}) or the Sydney University Molonglo Sky Survey (SUMSS, \citealt{Mauch2008}). More recently, they used a larger sample of confirmed blazars and radio data from the Faint Images of the Radio Sky at Twenty-cm survey (FIRST, \citealt{B&W1994}) to recognize additional blazar candidates (WIBRaLS2; \citealt{D'Abrusco2019}). They also established a second catalog, KDEBLLACS, which employs only W1-W2 and W2-W3 colors (i.e., excluding a W4 requirement) to detect fainter candidates. KDEBLLACS sources also must have a radio counterpart. The total number of blazar candidates from WIBRaLS, WIBRaLS2, and KDEBLLACS is $17,996$. A recent study by \cite{Menezes2019} analyzed the Sloan Digital Sky Survey spectra for available objects in the \cite{D'Abrusco2019} catalogs and estimated that $\gtrsim 40\%$ of them are contaminants, primarily quasars and galaxies. While spectroscopic follow-up is essential for determining the nature of blazar candidates, an approach for removing contaminants from a large pool of blazar candidates would greatly aid in more efficient follow-up investigations to determine their true nature. We notice that at the angular resolution of radio surveys used by \citealt{D'Abrusco2014,D'Abrusco2019}  ($\sim$45$^{\prime\prime}$ for both NVSS and SUMSS, $\sim$5$^{\prime\prime}$  for FIRST), the radio structure for most blazar candidates is not resolved, particularly for sources outside the FIRST footprint. 

In this work, we take advantage of the radio images of Very Large Array Sky Survey (VLASS, \citealt{Lacy2020, Gordon2021}). At the time of this study, VLASS provides two-epoch\footnote{In January 2023, VLASS started observing a third epoch of the sky. We only use the first two epochs in this study.} 3\,GHz images for all the sky above Decl.$>-40\,\deg$ at a higher angular resolution than previous large sky radio surveys ($\sim$2.5$^{\prime\prime}$). Since the blazar's jet is directed towards us at a small angle, the VLASS radio morphology is expected to be compact. 
We note that the actual morphology of blazars could be more complicated. 
For example, \cite{Kharb2010} showed that some of the blazars in the MOJAVE (Monitoring Of Jets in Active galactic nuclei with VLBA Experiments; \citealt{Lister2021}) sample exhibit both compact and extended emission at 1.4\,GHz.  However, we will show that at the VLASS frequency, resolution, and sensitivity, most of the MOJAVE blazars are classified as compact objects. Thus, if the VLASS images resolve a clear two-sided jet pattern, we can conclude that the radio source is likely not a blazar. 

The focus of this paper is the introduction of an automated algorithm that utilizes publicly available VLASS 3\,GHz radio images to classify sources whose morphology are compatible or not with being blazars. 
 This paper is structured as follows: In Section \ref{sec:data}, we present an overview of the data used in this study. Section \ref{sec:method} describes our algorithm for automatically classifying VLASS sources based on their morphology. We present the automated classification of all sources in Section \ref{sec:result}. In Section \ref{sec:discuss}, we discuss sources with mismatched positions in the blazar catalog. We also discuss appearances of MOJAVE blazars and their classification with VLASS 3\,GHz images. We summarize our work in Section \ref{sec:summ}. Throughout this study, we adopt a flat $\Lambda$CDM cosmology with $\mathrm{H}_{{\mathrm{0}}}$=70\,km~s$^{\mathrm{-1}}$~Mpc$^{\mathrm{-1}}$, $\Omega_{\mathrm{M}}$=0.3, and $\Omega_{\mathrm{\Lambda}}$=0.7, and all magnitudes are reported in the AB magnitude system.

\section{Data}
\label{sec:data}

Our dataset comprises a combination of three catalogs: Roma-BzCAT \citep{Massaro2015}, WIBRaLS \citep{D'Abrusco2014, D'Abrusco2019}, and KDEBLLACS \citep{D'Abrusco2019}. WIBRaLS and KDEBLLACS contain blazar candidates, while Roma-BzCAT consists of confirmed blazars. We download all available epochs of the VLASS Quick Look Images for the sources in this dataset. VLASS1.1 corresponds to the first epoch of the first half of the sky, and VLASS1.2 is the first epoch of the second part of the sky. The first epoch is completed in 2019, while the second epoch, VLASS2.1 and VLASS2.2, is completed in June 2022. There are upgrades in the image processing pipeline between epochs 1 and 2, producing cleaner radio images for both faint and bright sources by fixed pipeline issues (for more details, refer to \citealt{Lacy2022note13}).

The blazar candidates in WIBRaLS/WIBRaLS2 (hereafter referred to as WIBRaLS) are selected based on two criteria that must be satisfied simultaneously: i) having similar WISE colors (W1$-$W2 and W2$-$W3, and W3$-$W4) as known blazars in Roma-BzCat, and ii) having a radio counterpart. 
Meanwhile, KDEBLLACS complements WIBRaLS by adding fainter candidates not detected in the W4 band.  

Initially, our dataset includes a total of $12,416$ blazar candidates from the WIBRaLS and WIBRaLS2 catalogs. We find that $9,821$ candidates have VLASS Quick Look Image coverage, with $9,061$ sources having two VLASS epochs and $220$ sources having only one epoch. For KDEBLLACS, there are a total of $4,996$ out of $5,580$ sources with VLASS coverage, with $4,926$ two-epoch sources and $70$ one-epoch sources. Out of $3,561$ blazars in Roma-BzCAT, $3,134$ sources have VLASS images, with $3,078$ sources having two epoch images and $56$ sources having only one epoch. We do not differentiate between subclasses of blazars in this study.

\section{Method}
\label{sec:method}

\subsection{Automated algorithm of morphology recognition}
\label{subsec:algorithm}

In VLASS images, jet patterns are easily recognizable by the human eye, and manual morphological classification can be achieved through visual inspection. However, our objective is to develop an automated algorithm that can mimic the human visual classification in an objective and reproducible way. To achieve this, we download $2\arcmin\times2\arcmin$ VLASS images centered on known blazar/candidates with catalog coordinates, and utilize the \textit{zscale} algorithm \citep{Doug1986} to process the images into machine-readable masked values. The \textit{zscale} algorithm is specifically designed to display images near their median in a computationally efficient way without the need of computing the entire pixel value distribution histogram. This algorithm is particularly useful in highlighting peaked patterns that are brighter than the background, making them more discernible in astronomical images. The \textit{zscale} algorithm returns two limits in pixel values denoted as $zs_{up}$ and $zs_{low}$. Pixels whose values are above $zs_{up}$ appear bright while those with values below $zs_{low}$ appear dark.

\begin{figure*}{\includegraphics[scale=0.20]{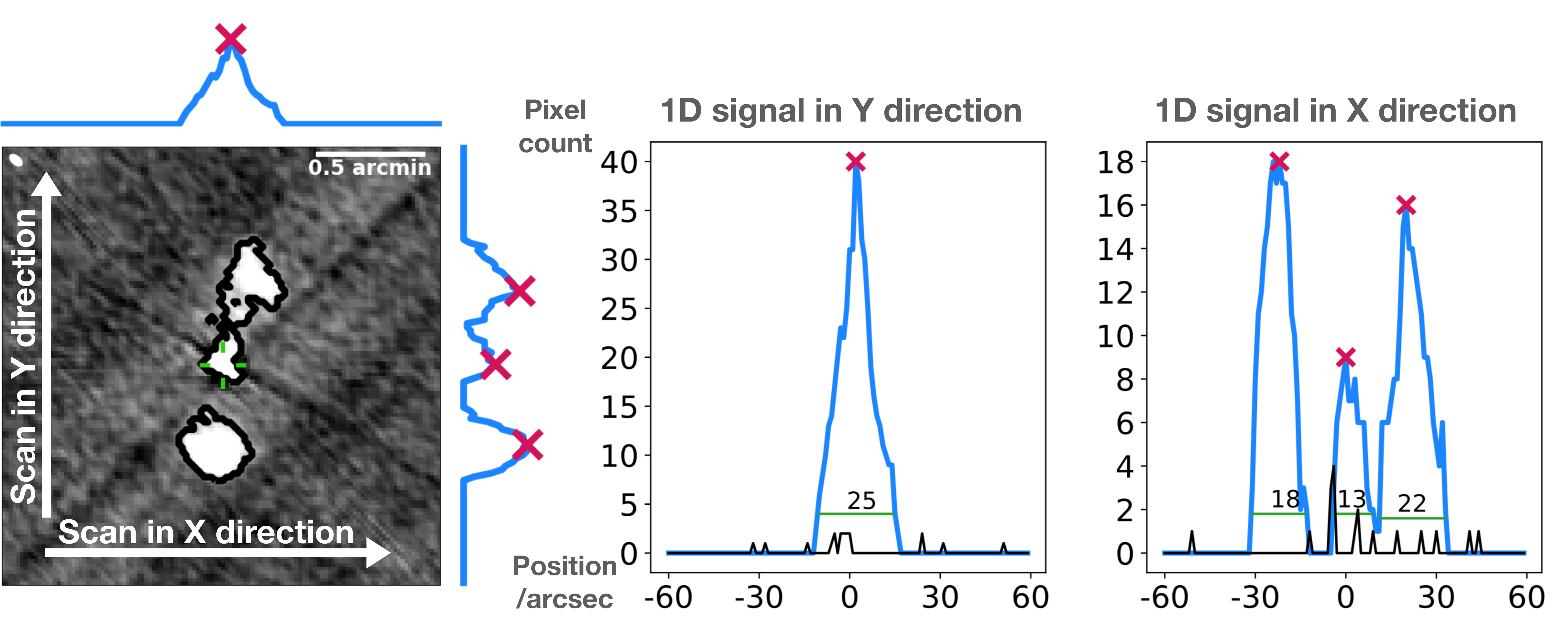}}
  \caption{An illustration of the automated morphological classification algorithm applied to VLASS images (The beam size is denoted in the top left corner, is approximately 1.8$^{\prime\prime}$$\times$2.9$^{\prime\prime}$). The algorithm processes each image, converting it into binary values (marked sections are depicted by black contour lines) based on their flux. Each of the 1D curve (blue line) represents the pixel count in a single scanning row in the selected direction, denoted by the white arrows in the image. Cross markers overlayed on the 1D curve indicate peak locations within the 1D curve, which are used to assign morphological classifications to the sources. A green line on the plot represents the peak width measured by each of their relative height. The value in number of pixels is shown, which also contributes to determining the morphological class of the sources (refer to Sec.~\ref{subsec:class} for details).
  }
  \label{fig:algorithm}
\end{figure*}

Next, our algorithm focuses specifically on pixels with values $\geq zs_{up}$, as these pixels provide a good representation of the source's morphology of their radio emission. By only calculating the distribution of the brightest pixels, the algorithm can obtain a proxy of the source's morphology, as shown by the black contours in Fig.~\ref{fig:algorithm}. 

The next goal is to classify the morphology, estimated by the highlighted pixel in each image. Our algorithm transforms the shape of the source in the VLASS image onto two 1D lines, generated by stacking the pixel counts in the respective direction. The height and width of the single-/multi-peaked 1D curve in the resulting plots represent the morphological characteristics of the radio source (see Fig.~\ref{fig:algorithm} for an example).

\begin{figure*}
  \centering
  \makebox[\textwidth][c]{\includegraphics[scale=0.50]{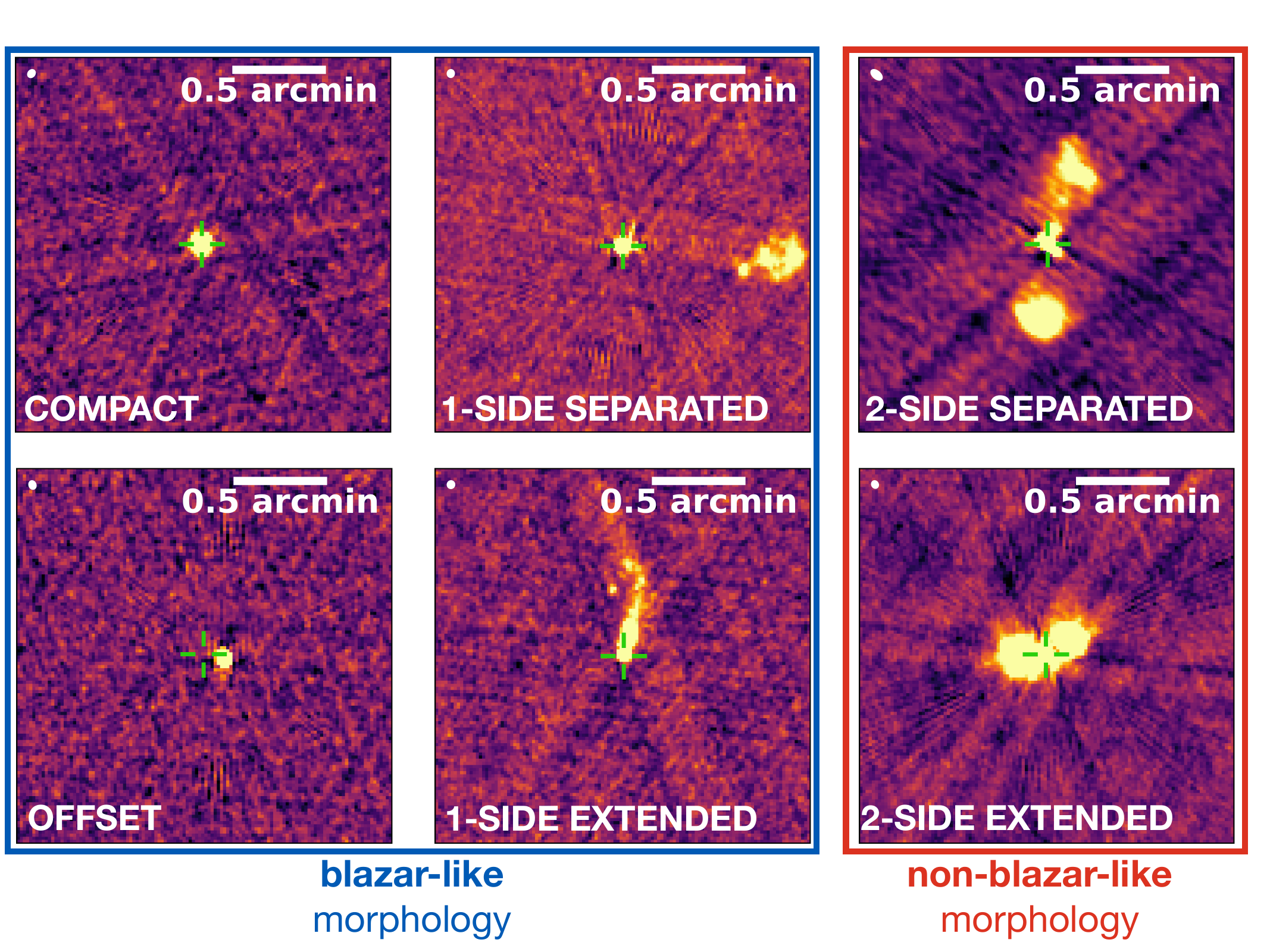}}
  \caption{Morphological classification of VLASS images using our automated algorithm. The images are categorized into six distinct morphological classes, grouped into two sets. Blazar-like morphologies include \textit{COMPACT}, \textit{OFFSET}, \textit{1-SIDE EXTENDED}, and \textit{1-SIDE SEPARATED}; while non-blazar-like morphologies comprise \textit{2-SIDE EXTENDED} and \textit{2-SIDE SEPARATED}. In each morphological class, their corresponding 1D signals are shown in two directions, indicated by blue lines. Automatically identified peaks are marked by pink crosses, with morphological classification determined by the properties of peaks, including number, width and distance to the center position. Descriptions of each morphological class and their characteristics can be found in Sec.~\ref{subsec:class}.}
  \label{fig:all_signal}
\end{figure*}

\begin{figure*}
  \centering
  \makebox[\textwidth][c]{\includegraphics[scale=0.50]{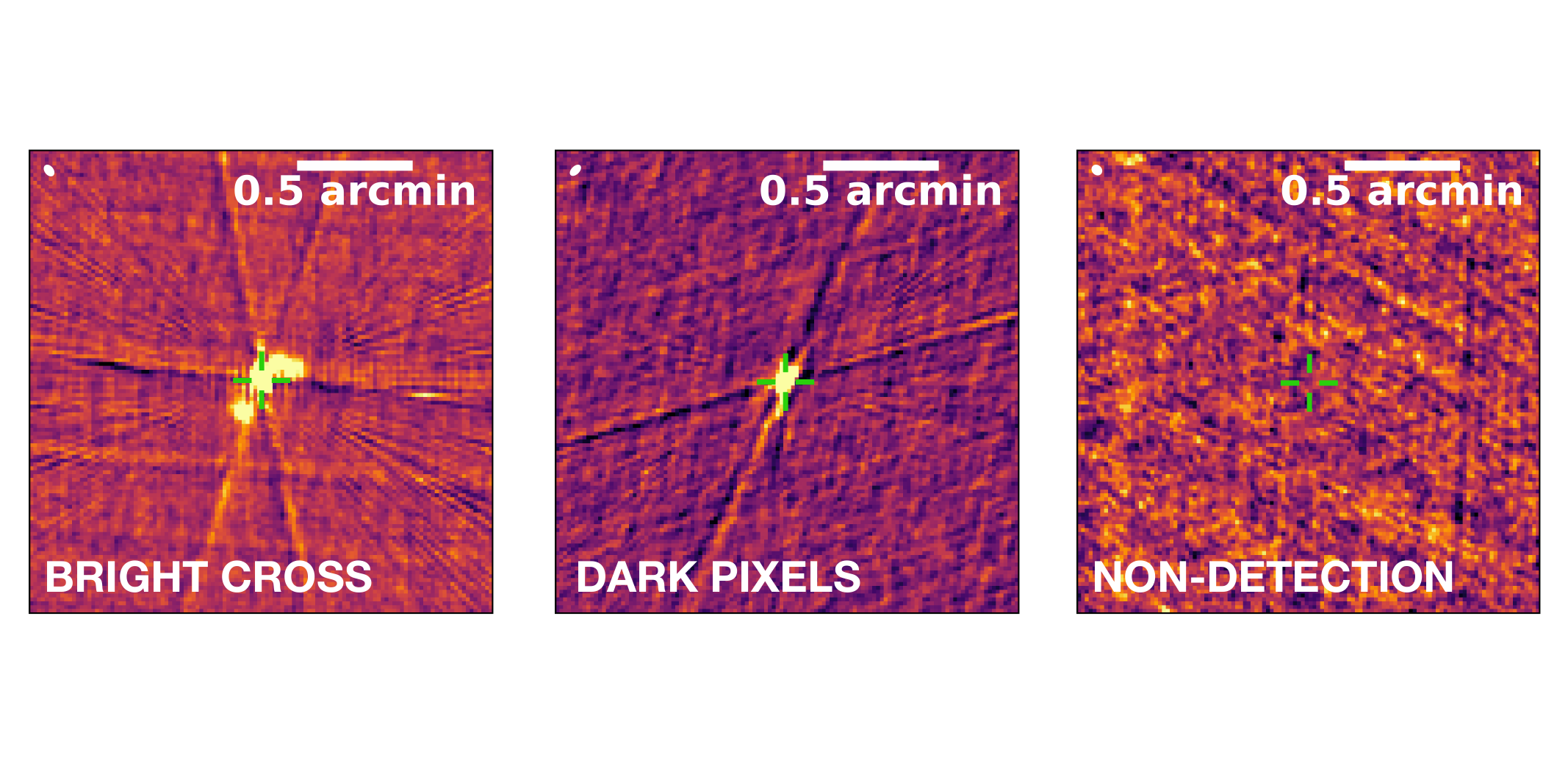}}
  \caption{An illustration of two artifact types and a \textit{NON-DETECTION} source in VLASS images. The algorithm identifies cross pattern and dark pixel artifacts, assigning appropriate quality flag values (1 for cross pattern presence, 2 for dark pixel presence, and 3 for both artifacts in the image). Sources with low signal-to-noise ratios that are difficult to discern by visual inspection are classified as \textit{NON-DETECTION} sources.}
  \label{fig:artifacts}
\end{figure*}

\subsection{Classifications on morphologies}
\label{subsec:class}

Our algorithm provides a morphological classification for each image by analyzing the shape of the signals obtained from processing the VLASS images into 1D curves. The algorithm classification can then be split into two groups: those displaying  blazar-like and non-blazar-like morphologies.
Examples of all subclasses are shown in Fig.~\ref{fig:all_signal}. Below, we explain how we identify each morphology classification from their 1D signal.

We categorize all sources with blazar-like morphologies into four subclasses: \textit{COMPACT}, \textit{OFFSET}, \textit{1-SIDE SEPARATED} and \textit{1-SIDE EXTENDED}. We list the details for each blazar-compatible subclass and their morphological features below:  

\begin{description}

    \item[\textit{COMPACT}]
        The VLASS image exhibits a compact unresolved source, indicating that the morphology is consistent with a jet directed towards us at a very small angle (see Fig.~\ref{fig:all_signal}, top left). A \textit{COMPACT} source is identified as a single peak in both row and column 1D signal, representing a single bright pixel cluster located in the central region of the VLASS image.
        
        \item[\textit{OFFSET}]
        The radio image shows one source that resembles the \textit{COMPACT} class, but whose optical/mid-infrared coordinates are offset from the compact radio source. To determine a source as \textit{OFFSET}, we examine the positions of two peaks in the row and column signals and calculate the offset value by measuring the distance to the catalog position. We classify sources as \textit{OFFSET} when their compact radio emission is at least 5$^{\prime\prime}$ away from the position listed in the catalogs (see Fig.~\ref{fig:all_signal}, bottom left as an example). This can happen when using lower-resolution radio surveys than VLASS and incorrectly associating the radio emission from another close source with its optical counterpart. A source is not classified as \textit{OFFSET} if its offset surpasses 10$^{\prime\prime}$, to avoid potential misidentification with an incorrect source deviating from the catalog position. For additional details, refer to Sec.~\ref{subsec:offset}.
        
    \item[\textit{1-SIDE SEPARATED}]
        The radio image reveals two distinct radio sources, with one precisely at the location of a confirmed blazar or blazar candidate. 
        Such a morphology is recognized as a typical blazar morphology in Roma-BzCAT (Fig.~\ref{fig:all_signal}, top middle). For a \textit{1-SIDE SEPARATED} source, the 1D signal displays exactly two peaks, either in row or column, or both, depending on the jet's orientation. If the jet aligns with the row or column direction, only a single peak will be observed in the corresponding direction.
        
    \item[\textit{1-SIDE EXTENDED}]
        The radio image exhibits a connected, extended structure with its core aligning with the catalog position of the blazar or blazar candidate. In this scenario, the jet extends to one side. Similar to the \textit{1-SIDE SEPARATED} morphology, this configuration also represents a central source emitting a jet oriented towards us at a small angle, and it is thus considered to be compatible with a blazar morphology (Fig.~\ref{fig:all_signal}, bottom middle). The 1D signal displays a single peak in both row and column, as the center is connected with the extended radio lobe, suggesting only one radio blob exists in the image. However, the peak should also have a width larger than 15$^{\prime\prime}$. We choose this value to be greater than a circle centering at the source, where we estimate the dynamic range to identify potential artifacts. This area has the radius of 7$^{\prime\prime}$ ($\sim$2-3 times of the beam size, see also Sec.~\ref{subsec:quality} for details). Consequently, the peak position is not centrally located but shifted in the projected direction that the jet extends.

\end{description}

We identify sources that show 2-side symmetrical jets as likely non-blazars, implying their viewing angle is likely not close to our line of sight. 
The two subclasses of sources with non-blazar-like morphologies are described below (also refer to the right side of Fig.~\ref{fig:all_signal}):

\begin{description}

    \item[\textit{2-SIDE SEPARATED}]
        The radio image displays two symmetrical jets on either side of the central core. 
        The observed comparable separations for both jets in \textit{2-SIDE SEPARATED} sources suggest that the jets are nearly perpendicular to our line of sight, thereby excluding them as blazars. The 1D signal exhibits exactly three peaks in row and/or column. Similar to the \textit{1-SIDE SEPARATED} morphology, the jet angle can influence the number of detected peaks. If the jets align with either row or column, fewer than three peaks will be displayed. However, the other direction will still identify three peaks, classifying the source as a \textit{2-SIDE SEPARATED} morphology.
        
    \item[\textit{2-SIDE EXTENDED}]
        The radio image presents two symmetrical jets, similar to the \textit{2-SIDE SEPARATED} objects. The distinguishing feature of \textit{2-SIDE EXTENDED} sources is that the lobes and the core are interconnected at the resolution of our images, exhibiting similar extension sizes. Following the same logic as the \textit{2-SIDE SEPARATED} case, these jets must be nearly perpendicular to our line of sight. In the 1D signal of a \textit{2-SIDE EXTENDED} source, the peak is closer to the center than in \textit{1-SIDE EXTENDED}, as the jets extending from the central source in both directions rather than just one. We determine that if the peak (center of the radio source) is within 10 $^{\prime\prime}$ ($\sim$3 times of the beam size, also being out of the 7$^{\prime\prime}$ circle where we assess the dynamical range) from the catalog position, it will be categorized as \textit{2-SIDE EXTENDED}; otherwise, it will be regarded as \textit{1-SIDE EXTENDED}.

\end{description} 

In addition to blazar-like and non-blazar-like classifications, we observe sources that exhibit no visible radio emission in VLASS images. We categorize these sources as \textit{NON-DETECTION}.

\begin{description}
    \item[\textit{NON-DETECTION}]
        The radio image displays no visible features near the catalog position. We assess the signal-to-noise ratio (S/N) to determine if the source is detected. The S/N is computed as the peak flux density within a 10$^{\prime\prime}$ circle surrounding the source, divided by the noise. The noise for each image is the root-mean-square value of all pixels, following a $2.5\sigma$ clipping of the image. If a source's S/N is below 5.0, it will be classified as \textit{NON-DETECTION}. The 1D signal reveals no significant peaks exceeding the noise level (for example, see the right panel in Fig.~\ref{fig:artifacts}).
\end{description}


This classification approach effectively distinguishes morphologies within a vast amount of radio data. The conversion from 2D images to 1D curves is both straightforward and rapid. The algorithm reduces the data processing complexity from $O(n^2)$ (pixels in two dimensions) to $O(n)$ (pixels in one dimension). Additionally, our classification method requires no prior models, calibrators, or training sets.

\subsection{Quality check}
\label{subsec:quality}

The majority of sources in all datasets are present in both VLASS epochs (VLASS 1.1/1.2 and VLASS 2.1/2.2, refer to Sec.~\ref{sec:data}). We independently apply the algorithm to these two epochs. We first assign quality flags to the image in each of the epoch. If both epochs result in the same morphological classification, it is considered final. However, in cases of discrepancy, the final classification will be determined by the image of superior quality. The process is as follows:

\begin{itemize}
    \item We first define a quality flag with a value ranging from 0 (best quality) to 3 (worst quality). Further details are provided below. If one of the epochs has a superior quality flag compared to the other, the morphology from that epoch is considered the final classification. If the quality flags are identical, but the classifications from each epoch remain inconsistent, we proceed to the next step.
    \item The algorithm examines artifacts in the images, which may result from deficits in the clean algorithm and/or inadequate phase calibration during image production and that can affect the morphological classification \citep{Lacy2022note13}. We employ a quality flag to assess the impact of artifacts on the image. There are two types of artifacts we pay attention to and that we referred to as `bright cross' and `dark pixels':
    
    \begin{enumerate}
    
    \item Bright cross (Fig.~\ref{fig:artifacts} left): This is a pattern of straight lines, composed of bright pixels (pixel value higher than $zs_{up}$) traversing the image. It is identified by detecting narrow, low peak signals extending across the image. In the 1D signal, the algorithm counts the number of all peaks with pixel count (height of the 1D signal) less than 2. If the image displays more than 10 low peaks in the row and column directions combined, the quality flag is incremented by 1.
    
    \item Dark pixels (Fig.~\ref{fig:artifacts} middle): This pattern consists of dark pixels (pixel value lower than $zs_{low}$) primarily located adjacent to the central bright source. The algorithm detects all pixels with values below $zs_{low}$ in the 1D signal. After visually inspecting several instances of this artifact, we establish that if the sum of all dark pixels exceeds 30\% of the sum of all bright pixels, the quality flag is incremented by 2.

    \end{enumerate}
    
\end{itemize}
    
    If one epoch has a lower-quality flag value, the morphology of that epoch is considered final. In cases where both images have the same quality flag, thus similarly affected by artifacts, we proceed to the following step.
    
    \begin{itemize}
    \item The algorithm calculates the ratio of the total sum of dark pixels to the total sum of bright pixels. The epoch with the smaller ratio prevails and is chosen as the final classification.
    \end{itemize}
    
    In addition to the aforementioned procedure, we implement an extra check for sources to be classified as \textit{COMPACT} by assessing the dynamic range value. This is because artifacts can mislead the algorithm into incorrectly interpreting them as sidelobes. The dynamic range is a metric to discern the nature of a point source, mitigating the effects of artifact presence. To quantify the dynamic range, we employ the `peak-to-ring' metric from \citealt{Gordon2021} and \citealt{Lacy2022note13}. We first calculate the peak flux value within a 2$\arcsec$ circle centered on the catalog position of the source, and then measure another peak flux within a 1$\arcsec$ wide annulus, where its inner boundary is 7$\arcsec$ away from the center (approximately 2 times the beam size for VLASS). We apply this measure to distinguish a core-dominated source when the images are affected by artifacts (when the epoch with the smallest quality flag value is greater than 0). In this case, if in any epoch the `peak-to-ring' ratio is larger than 2, the source will be classified as a \textit{COMPACT} source. If none of the epochs (including the single-epoch sources) exceeds 2 in the `peak-to-ring' ratio, the source will be assigned as \textit{VISUAL NEEDED} for further check. The final classification of these sources will be determined and assigned manually by us, following a visual check.
    
    We assign a special category \textit{COMPACT\_D} to sources that comply with all the following: (i) a `peak-to-ring' dynamic range greater than two, (ii) the smallest quality flag between both epochs exceeds one, and (iii) the source was classified as non-\textit{COMPACT} in both VLASS epochs. This implies that the artifacts probably cause the seemingly extended feature. For the sources with only one epoch image available, the source will first be checked if it is detected. It will be classified as a \textit{NON-DETECTION} if its S/N value is below 5.0. The automated classification is final when the quality flag is not larger than one. If the quality flag of the image exceeds one, and the `peak-to-ring' dynamic range is greater than two, the source will be assigned as \textit{COMPACT\_D}. Otherwise, the source will be classified as \textit{VISUAL NEEDED} to be visually classified.

\section{Result of the algorithm classification}
\label{sec:result}

We apply our morphological classification algorithm to a total of $9,821$ VLASS sources from the WIBRaLS catalog, $4,996$ sources from the KDEBLLACS catalog, and $3,134$ sources from the confirmed blazars in the Roma-BzCAT catalog. A portion of the results for the Roma-BzCAT catalog is presented in Table \ref{tab:sample_catal_roma}. This table includes the name generated by the algorithm in the form of Jhh:mm:ss.ss$\pm$dd:mm:ss.ss, coordinates, and the assigned morphological classification. Blazar-like morphologies consist of \textit{COMPACT}, \textit{1-SIDE SEPARATED}, and \textit{1-SIDE EXTENDED} categories. Non-blazar-like morphologies are \textit{2-SIDE SEPARATED} and \textit{2-SIDE EXTENDED}. 

In Roma-BzCAT, we conduct a visual inspection of all non-blazar-like sources. In cases where the visual classification differs from the algorithm-based classification, we assign the accurate classification to `VClass.' We discuss and show VLASS 2 images of all non-blazar-like sources in the Appendix. The tables with the same entries have also been generated for the WIBRaLS and KDEBLLACS catalogs (refer to Table \ref{tab:sample_catal_wibrals} for WIBRaLS and Table \ref{tab:sample_catal_kdebllacs} for KDEBLLACS). We do not visually check all non-blazar-like sources in these two catalogs.

\begin{table*}
  \renewcommand{\arraystretch}{0.8}
  \begin{center}
  \setlength{\tabcolsep}{0.014\linewidth}
  \begin{tabular}{l l l l l l l l}
  \hline
  \hline
  \noalign{\smallskip} No.$^{a}$ & Name$^{b}$ & R.A.$^{c}$ & Decl.$^{d}$& Class$^{e}$ & VFlag$^{f}$ & VClass$^{g}$ & SEpoch$^{h}$\\ 
  \hline
  \noalign{\smallskip}
    1 & J000020.39-322101.00 & 0.08495833333 & -32.35027778 & COMPACT\_D & 0 & &\\
    2 & J000105.29-155106.98 & 0.2720416667 & -15.85193889 & COMPACT & 0 & &\\
    3 & J000108.62+191434.18 & 0.2859166667 & 19.24282778 & COMPACT & 0 & &\\
    156	& J010838.76+013500.31 & 17.1615    & 1.583419444 & COMPACT & 0	& & EP02\\
    1185 & J100021.79+223318.61 & 150.0907917 & 22.55516944 & 2-SIDE EXTENDED & 1 & & \\
    2566 & J174805.82+340401.20 & 267.02425 & 34.067 & 2-SIDE EXTENDED & 1 & COMPACT &\\
    2677 & J193109.60+093717.50	& 292.7882083$^{*}$ & 9.6211944$^{*}$ & OFFSET & 1 & &\\
    2805 & J211720.72+050257.58 & 319.3363333 & 5.049327778 & VISUAL NEEDED & 1 & COMPACT & \\
    3132 & J235846.08+195520.31 & 359.692 & 19.92230833 & COMPACT\_D & 0 & &\\
    3133 & J235859.86+392228.30 & 359.7494167 & 39.37452778 & COMPACT\_D & 0 & &\\
    3134 & J235933.18+385042.28 & 359.88825 & 38.84507778 & COMPACT & 0 & & \\
  \noalign{\smallskip}
  \hline
  \hline
  \noalign{\smallskip}
  \end{tabular}
  \tablecomments{$^{(a)}$: Number in the result catalog; $^{(b)}$: Name of the source; $^{(c)}$: Right Ascension (J2000); $^{(d)}$: Declination (J2000); $^{(e)}$: Morphological classification decided by the algorithm; $^{(f)}$: Visual Flag, 1 if the source is visually inspected, 0 if not; $^{(g)}$: Visual classification, assigned by human eyes if the algorithm classification is wrong, this entry is left empty if the source is not visually inspected or the algorithm classification is correct when `VFlag’ is 1; $^{h}$: If the source has two epochs available, this column is left empty, otherwise it will denote which epoch of VLASS images is available for the source, EP01: VLASS 1, EP02: VLASS 2; \\$^*$: Corrected value by cross-matching with MilliQuas catalog, more details refer to Sec.~\ref{subsec:offset}.}
  \caption{Sources from the Roma-BzCAT catalog processed and classified using our algorithm. Each source is analyzed based on available radio images from two VLASS epochs, with the algorithm assigning a final morphological classification. Each visually inspected source has a `VFlag' of 1. If the visual classification (`VClass') does not match the algorithm classification (`Class'), we update `VClass' to the correct classification. For sources labeled with \textit{VISUAL NEEDED}, their class is determined visually and assigned to `VClass'.} We look into all non-blazar-like sources according to the algorithm in Roma-BzCAT to verify their morphology; more details are in the Appendix. The full table is available online.
  \label{tab:sample_catal_roma}
  \end{center}
\end{table*}

\begin{table*}
  \renewcommand{\arraystretch}{0.8}
  \begin{center}
  \setlength{\tabcolsep}{0.014\linewidth}
  \begin{tabular}{l l l l l l l l}
  \hline
  \hline
  \noalign{\smallskip} No. & Name & R.A. & Decl.& Class & VFlag & VClass & SEpoch\\ 
  \hline
  \noalign{\smallskip}
1 & J000020.40-322101.24 & 0.085 & -32.35034444 & COMPACT\_D & 0 &\\
2 & J000029.08-163620.24 & 0.1211666667 & -16.60562222 & COMPACT & 0 &\\
3 & J000047.05+312028.21 & 0.1960416667 & 31.34116944 & COMPACT & 0 &\\
4 & J000101.05+240842.52 & 0.254375 & 24.14514444 & COMPACT & 0 &\\
5 & J000105.29-155107.21 & 0.2720416667 & -15.85200278 & COMPACT & 0 &\\
9817 & J235919.53-204756.10 & 359.831375 & -20.79891667 & 2-SIDE EXTENDED & 0 &\\
9818 & J235931.80-063943.37 & 359.8825 & -6.662047222 & COMPACT & 0 &\\
9819 & J235935.23+522236.85 & 359.8967917 & 52.37690278 & COMPACT & 0 &\\
9820 & J235941.29+392439.47 & 359.9220417 & 39.41096389 & COMPACT & 0 &\\
9821 & J235951.04+470709.41 & 359.9626667 & 47.11928056 & COMPACT & 0 &\\
  \noalign{\smallskip}
  \hline
  \hline
  \end{tabular}
  \caption{Same as Tab.~\ref{tab:sample_catal_roma} for WIBRaLS.}
  \label{tab:sample_catal_wibrals}
  \end{center}
\end{table*}

\begin{table*}
  \renewcommand{\arraystretch}{0.8}
  \begin{center}
  \setlength{\tabcolsep}{0.014\linewidth}
  \begin{tabular}{l l l l l l l l}
  \hline
  \hline
  \noalign{\smallskip} No. & Name & R.A. & Decl.& Class & VFlag & VClass & SEpoch\\ 
  \hline
  \noalign{\smallskip}
1 & J000007.63+420725.51 & 0.03179166667 & 42.12375278 & NON-DETECTION & 0 &\\
2 & J000010.29-363405.26 & 0.042875 & -36.56812778 & COMPACT & 0 &\\
3 & J000056.23-082742.05 & 0.2342916667 & -8.461680556 & COMPACT & 0 &\\
4 & J000116.38+293534.59 & 0.31825 & 29.59294167 & COMPACT & 0 &\\
5 & J000126.44+733042.60 & 0.3601666667 & 73.51183333 & COMPACT & 0 &\\
4992 & J235859.76+431617.64 & 359.749 & 43.27156667 & COMPACT & 0 &\\
4993 & J235901.15+171925.84 & 359.7547917 & 17.32384444 & COMPACT & 0 &\\
4994 & J235932.16-121022.56 & 359.884 & -12.17293333 & COMPACT & 0 &\\
4995 & J235944.89+054431.34 & 359.9370417 & 5.742038889 & COMPACT & 0 &\\
4996 & J235955.31+314559.85 & 359.9804583 & 31.766625 & COMPACT & 0 &\\
  \noalign{\smallskip}
  \hline
  \hline
  \end{tabular}
  \caption{Same as Tab.~\ref{tab:sample_catal_roma} for KDEBLLACS.}
  \label{tab:sample_catal_kdebllacs}
  \end{center}
\end{table*}

Table~\ref{tab:result_cat} provides a summary of our morphological classification of all sources from the Roma-BzCAT, WIBRaLS, and KDEBLLACS catalogs with corresponding VLASS images. Approximately 95\% of sources in the Roma-BzCAT catalog exhibit morphologies consistent with blazars, while 86\% of sources in the WIBRaLS catalog and 88\% of sources in the KDEBLLACS catalog are similarly consistent with being blazars.

The lower ``contamination" rate (i.e., sources unlikely to be blazars) within the Roma-BzCAT catalog is expected, as this curated catalog requires each source to have spectroscopic information to establish its blazar type (BL Lac or FSRQ). However, it is noteworthy that 4.5\% of sources (141) are classified as non-blazar-like sources. We visually inspect each source and find that 106 sources indeed show morphologies that are likely not blazars. A closer examination of the multi-wavelength properties of these 106 Roma-BzCAT sources is required to ascertain their actual nature. All sources are shown in the Appendix.

To evaluate the algorithm's accuracy, we randomly select 1000 sources from the Roma-BzCAT catalogs. We conduct a visual inspection of these 1000 sources and assign a visual morphology classification, following the same naming convention used by our algorithm. By considering the visual inspection classification as the ground truth, the accuracy of the classifier is at the percent of 90.6\% for the 1000 sources sample. The classifier makes mistakes by confusing artifacts with actual features in the morphology. These artifacts, as shown in Fig.~\ref{fig:artifacts}, have signals that resemble the real radio signals emitted by the source. 

\section{Discussion}
\label{sec:discuss}

\subsection{Misidentifications on \textit{OFFSET} sources}
\label{subsec:offset}

\textit{OFFSET} sources are present in all three catalogs of our dataset. This class of sources is characterized by the absence of radio emission at the catalog position. Instead, radio emission is observed with an offset ($5-10\arcsec$) from the position registered in the catalog. For \textit{OFFSET} sources in Roma-BzCAT, it is important to note that the spectroscopic information for the central source has already been obtained, allowing for the determination of a specific blazar type \citep{Massaro2009}. 

\begin{figure*}
  \centering
  \makebox[\textwidth][c]{\includegraphics[scale=0.10]{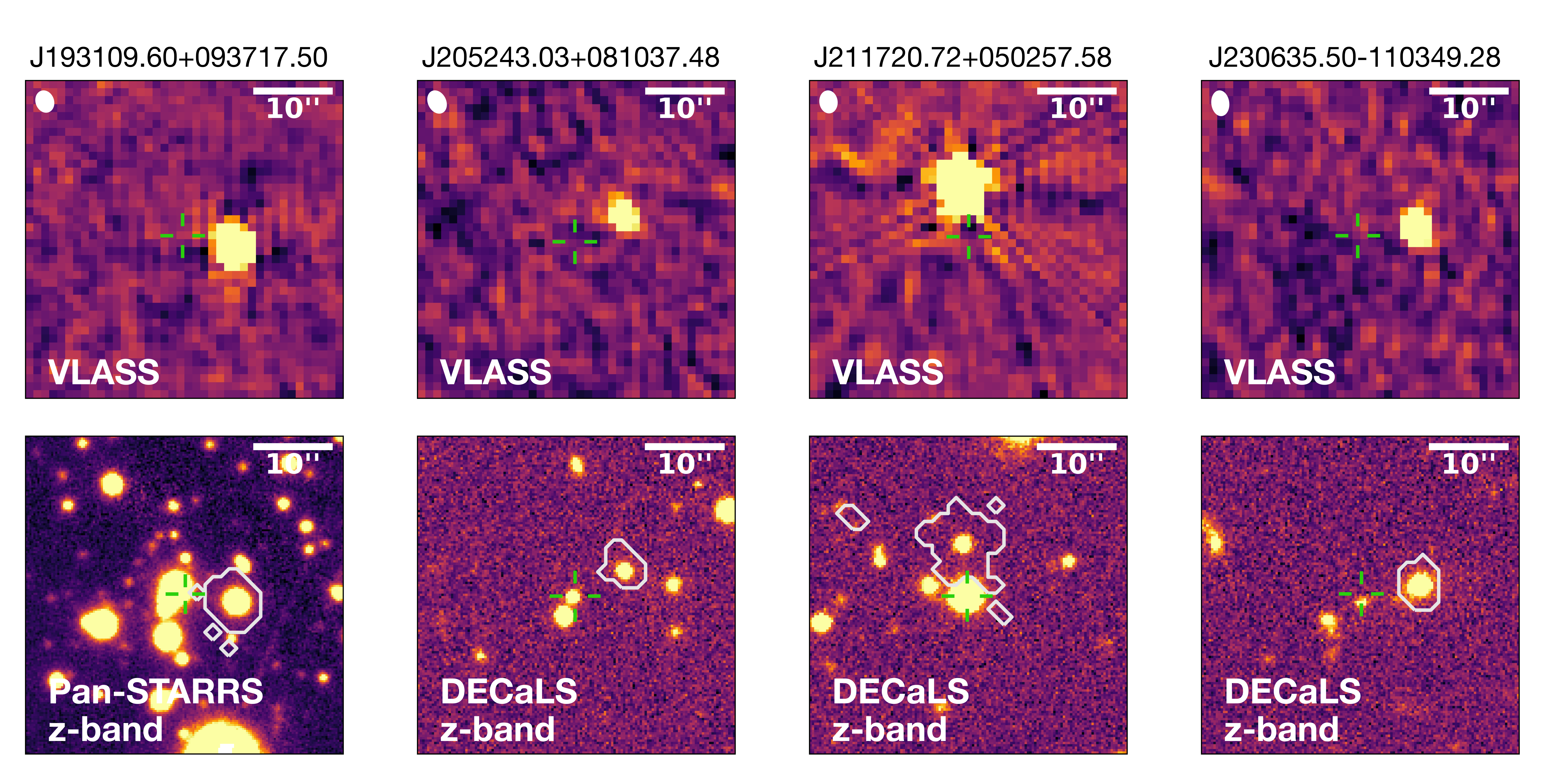}}
  \caption{\textit{OFFSET} sources from Roma-BzCAT, four in total. These sources are classified based on the discrepancy between the catalog position and the position of the radio source. The green cross represents the source position recorded in Roma-BzCAT. The radio image (upper row) demonstrates that the radio-emitting source is not located at the catalog position but has an offset of more than 5$^{\prime\prime}$. In the optical z-band image (lower row), a separate radio-silent source is found at the catalog position. This mismatch is a position error in Roma-BzCAT and we provide the accurate coordinates in our result.}
  \label{fig:offset}
\end{figure*}

For the six \textit{OFFSET} sources in the Roma-BzCAT catalog, we examine their available archival data. Four of the six sources show real offset (shown in Fig.~\ref{fig:offset}); the other two sources show no actual offset in their VLASS images and the optical counterpart is at the exact position of the radio source. Thus, they are classified as \textit{COMPACT} sources instead. We perform crossmatching to the MilliQuas catalog \citep{Flesch2021} within a range of 5--10$^{\prime\prime}$ and discover that all four sources were mispositioned in the Roma-BzCAT catalog; instead, all four sources have previous information on the precise position of their counterparts in X-ray or other wavelengths. The positions reported in the literature align with the actual VLASS radio source and there is no offset in their optical counterpart as well. Here, we provide information of the optical counterparts information for the four \textit{OFFSET} sources:

\textit{J193109.60+093717.50}: The optical position is provided in \cite{Motch1998}. They carried out the observation at Observatoire de Haute-Provence, CNRS, France, to acquire the optical position and also identify a featureless spectrum to confirm its BL Lac nature.

\textit{J205243.03+081037.48}: It is included in a spectroscopic compaign done by \cite{Piranomonte2007}. They observe this source with the 3.6m Telescopio Nazionale Galileo (TNG). The optical position is determined and its spectrum also shows the absence of emission or absorption features, a BL Lac characteristic.

\textit{J230635.50-110349.28} and \textit{J232352.50+421054.98}: Their positions are recorded in \cite{Bauer2000}. They carry out cross-identifications to match NVSS sources to bright X--ray sources and provide the coordinates of the closest matches. \textit{J232352.50+421054.98} was classified as a BL Lac object based on its archival spectrum.

We inspect all coordinates above and confirm that they are located on the center of radio sources in VLASS images. We have updated the positions of these mispositioned sources in the Roma-BzCAT catalog and included them in our output catalog for accurate reference (see Tab.~\ref{tab:sample_catal_roma} for an example).

\begin{table*}[!ht]
  \centering
  \setlength{\tabcolsep}{0.010\linewidth}
  \begin{tabular}{ l|| l | l l l l|l l|l}
  \multicolumn{2}{c}{ } & \multicolumn{4}{c}{\large blazar-like morphology} & \multicolumn{2}{c}{\large non-blazar-like morph.} \\
  \hline
  \hline
    Catalog & TOTAL & COMPACT/\_D & COMPACT & 1-SIDE & 1-SIDE  & 2-SIDE & 2-SIDE & NON-  \\

   &    &   & OFFSET & SEP & EXT & SEP & EXT & DET \\ \hline
  Roma-BzCAT & 3134 & 2835/1144 (90.5\%) & 6 (<1\%) & 112 (3.6\%) & 20 (<1\%) & 36$^{*}$ (1.1\%) & 105$^{*}$ (3.4\%) & 13 (<1\%)\\
  WIBRaLS & 9821 & 7326/1491 (74.6\%) & 72 (<1\%) & 958 (9.8\%) & 64 (<1\%) & 703 (7.2\%) & 506 (5.2\%) & 90 (<1\%)\\
  KDEBLLACS & 4996 & 4103/67 (82.1\%) & 233 (4.7\%) & 76 (1.5\%) & 6 (<1\%) & 36 (<1\%) & 3 (<1\%) & 356 (7.1\%)\\
  \hline
  \hline
  \end{tabular}
  \caption{Statistics of different VLASS morphological classifications for all sources in Roma-BzCAT, WIBRaLS, and KDEBLLACS. We identify four classes as compatible with blazar-like morphologies and two classes as non-blazar-like morphology. The sources that are classified as \textit{COMPACT\_D} are included in \textit{COMPACT} sources (see details in Sec.~\ref{subsec:quality}) and the separate number of this class is listed}. The percentage of non-blazars in WIBRaLS (12\%) is higher than in Roma-BzCAT (4\%), which is expected given that WIBRaLS consists of candidates, while Roma-BzCAT is a curated catalog. In a test sample of 1000 sources from Roma-BzCAT, the classification accuracy is 90.6\%.
  \tablecomments{$^*$: We conduct a visual examination of every non-blazar-like source in Roma-BzCat. The number reported here is the result after visual inspection, ruling out all false classifications. For more details, refer to the Appendix.}
  \label{tab:result_cat}
\end{table*}

\subsection{Testing the algorithm on the MOJAVE blazars}
\label{sec:mojave}

The MOJAVE survey focuses on obtaining high-resolution images of AGN radio jets using VLBA observations, providing detailed images at the parsec level. We ran our VLASS classifier algorithm in 320 MOJAVE sources that are identified as blazars \citep{Lister2021}. The result is that 298 are classified to have blazar-like morphology. The VLASS images of the 22 non-blaza-like sources (\textit{2-SIDE EXTENDED} or \textit{2-SIDE SEPARATED})are shown in Fig.~\ref{fig:mojave}. Out of those, 11 are still consistent with being compact sources, according to our visual inspection, resulting in 11 MOJAVE blazars with extended VLASS morphology. Our full classification for all 320 MOJAVE blazars is presented in an accompanying on-line table.

There are a few cases in which the MOJAVE blazars are known to display extended emission surrounding their compact core. For example, TXS 0716+714 \citep[][]{Antonucci1986, Wagner1996},  PKS 1045--18  and  PKS 1036+054 (see Fig.~2 in \citealt{Kharb2010}). In these three examples, our algorithm classifies the sources as  \textit{COMPACT\_D}. This means that some \textit{COMPACT\_D} sources might actually be extended (see Sec.~\ref{subsec:quality}), but still core-dominated.

We find that 15 out of 320 MOJAVE sources are classified under \textit{COMPACT\_D}, including the examples above. However, without the knowledge of additional radio images of higher resolution or at different frequencies, we can not differentiate if the extended morphology is real or introduced by artifacts.

In summary, MOJAVE blazars can display extended structures at both the submilliarcsecond-scale in VLBA images and the kiloparsec-scale in VLA 1.4\,GHz images. However, in most cases, these extended features are not discernible in the frequency and sensitivity of VLASS 3\,GHz images due to low resolution or sensitivity. As a result, most sources are categorized as \textit{COMPACT} objects, in line with their blazar characteristics (but see the Appendix for discussion and images of MOJAVE sources classified as non-COMPACT).

\begin{table*}
  \renewcommand{\arraystretch}{0.8}
  \begin{center}
  \setlength{\tabcolsep}{0.014\linewidth}
  \begin{tabular}{l l l l l l l l}
  \hline
  \hline
  \noalign{\smallskip} No. & Name & R.A. & Decl.& Class & VFlag & VClass & SEpoch\\ 
  \hline
  \noalign{\smallskip}
1 & S4 0003+38 & 1.48823 & 38.33754 & COMPACT & 0 & &\\ 
2 & NRAO 005 & 1.55789 & -6.39315 & COMPACT & 0 & &\\ 
3 & CRATES J0009+0628 & 2.26638 & 6.47257 & COMPACT & 0 & &\\ 
4 & III Zw 2 & 2.62919 & 10.97486 & COMPACT & 0 & &\\ 
5 & 4C +40.01 & 3.37971 & 40.86032 & 2-SIDE EXTENDED & 1 & &\\ 
405 & 1ES 2344+514 & 356.77015 & 51.70497 & COMPACT & 0 & &\\ 
406 & PKS 2345-16 & 357.01087 & -16.52001 & COMPACT\_D & 0 & &\\ 
407 & 4C +45.51 & 358.59033 & 45.88451 & COMPACT & 0 & &\\ 
408 & S5 2353+81 & 359.09497 & 81.88118 & COMPACT & 0 & &\\ 
409 & PKS 2356+196 & 359.69202 & 19.92231 & COMPACT & 0 & &\\ 
  \noalign{\smallskip}
  \hline
  \hline
  \end{tabular}
  \caption{Same as Tab.~\ref{tab:sample_catal_roma} for MOJAVE sources.}
  \label{tab:mojave}
  \end{center}
\end{table*}

\section{Summary}
\label{sec:summ}

We develop an algorithm to classify radio sources into blazar-like or non-blazar-like based on their radio morphology on VLASS images. The algorithm classifies all sources into six morphological classes: four that are consistent with blazar morphologies and two non-blazar-like morphologies (see Fig.~\ref{fig:all_signal} and Sec.~\ref{sec:method}). We apply the algorithm on three catalogs: WIBRaLS, which consists of blazar candidates with similar WISE color to known blazars; KDEBLLACS, which also includes blazar candidates selected by WISE color but undetected in W4-band; and Roma-BzCAT, a catalog of known blazars based on a combination of spectroscopic identification, X-ray luminosity, and radio intensity and morphology in previous radio surveys. Our results indicate that all catalogs have sources that are unlikely to be blazars based on their morphology. In particular, WIBRaLS has a higher percentage of non-blazars (14\%) compared to KDEBLLACS (12\%), while Roma-BzCAT has the lowest percentage of seemingly non-blazars (5\%, see Table~\ref{tab:result_cat}) based on VLASS morphology. This is likely due to the low resolution of the radio surveys used by Roma-BzCAT, WIBRaLS, and KDEBLLACS to acquire radio counterparts and check morphology. Further details of our findings can be found in Sec.~\ref{sec:result}. We provide the morphological classifications in Tables \ref{tab:sample_catal_roma} $-$ \ref{tab:sample_catal_kdebllacs}. The full tables will be available online in the journal.

We notice a specific population in all catalogs where the positions of some sources in the catalog do not match their radio source positions. We refer to these sources as \textit{OFFSET}. Based on our analysis of 4 \textit{OFFSET} sources in Roma-BzCAT, these discrepancies are due to positional errors. We provide the corrected position from cross-matching with other catalogs (see Sec.~\ref{subsec:offset}). 

We also ran our algorithm on MOJAVE blazars, a collection of VLBA-monitored radio-active AGNs (Sec.~\ref{sec:mojave}). Although some MOJAVE blazars are known to show extended emission in VLBA and 1.4\,GHz VLA images, 93\% of MOJAVE blazars are classified as blazar-like sources by our algorithm using the VLASS 3\,GHz images.


Our algorithm offers an efficient and rapid tool for identifying contaminants and cleaning blazar candidate catalogs. Further investigations involving spectroscopic and multiwavelength follow-up can aid in pinpointing the true blazars and contaminants, and determining their properties. 
The future survey from SKA-Mid\footnote{https://www.skao.int/index.php/en/science-users/118/ska-telescope-specifications} \citep{Dewdney2009} will have higher sensitivity and resolution than VLASS.
This will allow us to examine the morphologies of radio sources, such as blazars, even in their earliest stages or at higher redshifts than currently possible.

The codes to reproduce this work and the resulting tables are available on Zenodo \citep{zenedo} under an open-source 
Creative Commons Attribution license: \dataset[doi:10.5281/zenodo.10124636]{https://zenodo.org/doi/10.5281/zenodo.10124635}.

\section*{Acknowledgements}


We thank the referee for the critical assessment of the paper, pointing out caveats of this work, which led to a substantial improvement of the manuscript.

The National Radio Astronomy Observatory is a facility of the National Science Foundation operated under cooperative agreement by Associated Universities, Inc.

This publication makes use of data products from the Wide-field Infrared Survey Explorer, which is a joint project of the University of California, Los Angeles, and the Jet Propulsion Laboratory/California Institute of Technology, funded by the National Aeronautics and Space Administration.

The Legacy Surveys consist of three individual and complementary projects: the Dark Energy Camera Legacy Survey (DECaLS; Proposal ID \#2014B-0404; PIs: David Schlegel and Arjun Dey), the Beijing-Arizona Sky Survey (BASS; NOAO Prop. ID \#2015A-0801; PIs: Zhou Xu and Xiaohui Fan), and the Mayall z-band Legacy Survey (MzLS; Prop. ID \#2016A-0453; PI: Arjun Dey). DECaLS, BASS and MzLS together include data obtained, respectively, at the Blanco telescope, Cerro Tololo Inter-American Observatory, NSF’s NOIRLab; the Bok telescope, Steward Observatory, University of Arizona; and the Mayall telescope, Kitt Peak National Observatory, NOIRLab. Pipeline processing and analyses of the data were supported by NOIRLab and the Lawrence Berkeley National Laboratory (LBNL). The Legacy Surveys project is honored to be permitted to conduct astronomical research on Iolkam Du’ag (Kitt Peak), a mountain with particular significance to the Tohono O’odham Nation.

NOIRLab is operated by the Association of Universities for Research in Astronomy (AURA) under a cooperative agreement with the National Science Foundation. LBNL is managed by the Regents of the University of California under contract to the U.S. Department of Energy.

This project used data obtained with the Dark Energy Camera (DECam), which was constructed by the Dark Energy Survey (DES) collaboration. Funding for the DES Projects has been provided by the U.S. Department of Energy, the U.S. National Science Foundation, the Ministry of Science and Education of Spain, the Science and Technology Facilities Council of the United Kingdom, the Higher Education Funding Council for England, the National Center for Supercomputing Applications at the University of Illinois at Urbana-Champaign, the Kavli Institute of Cosmological Physics at the University of Chicago, Center for Cosmology and Astro-Particle Physics at the Ohio State University, the Mitchell Institute for Fundamental Physics and Astronomy at Texas A\&M University, Financiadora de Estudos e Projetos, Fundacao Carlos Chagas Filho de Amparo, Financiadora de Estudos e Projetos, Fundacao Carlos Chagas Filho de Amparo a Pesquisa do Estado do Rio de Janeiro, Conselho Nacional de Desenvolvimento Cientifico e Tecnologico and the Ministerio da Ciencia, Tecnologia e Inovacao, the Deutsche Forschungsgemeinschaft and the Collaborating Institutions in the Dark Energy Survey. The Collaborating Institutions are Argonne National Laboratory, the University of California at Santa Cruz, the University of Cambridge, Centro de Investigaciones Energeticas, Medioambientales y Tecnologicas-Madrid, the University of Chicago, University College London, the DES-Brazil Consortium, the University of Edinburgh, the Eidgenossische Technische Hochschule (ETH) Zurich, Fermi National Accelerator Laboratory, the University of Illinois at Urbana-Champaign, the Institut de Ciencies de l’Espai (IEEC/CSIC), the Institut de Fisica d’Altes Energies, Lawrence Berkeley National Laboratory, the Ludwig Maximilians Universitat Munchen and the associated Excellence Cluster Universe, the University of Michigan, NSF’s NOIRLab, the University of Nottingham, the Ohio State University, the University of Pennsylvania, the University of Portsmouth, SLAC National Accelerator Laboratory, Stanford University, the University of Sussex, and Texas A\&M University.

BASS is a key project of the Telescope Access Program (TAP), which has been funded by the National Astronomical Observatories of China, the Chinese Academy of Sciences (the Strategic Priority Research Program “The Emergence of Cosmological Structures” Grant \# XDB09000000), and the Special Fund for Astronomy from the Ministry of Finance. The BASS is also supported by the External Cooperation Program of Chinese Academy of Sciences (Grant \# 114A11KYSB20160057), and Chinese National Natural Science Foundation (Grant \# 12120101003, \# 11433005).

The Legacy Survey team makes use of data products from the Near-Earth Object Wide-field Infrared Survey Explorer (NEOWISE), which is a project of the Jet Propulsion Laboratory/California Institute of Technology. NEOWISE is funded by the National Aeronautics and Space Administration.

The Legacy Surveys imaging of the DESI footprint is supported by the Director, Office of Science, Office of High Energy Physics of the U.S. Department of Energy under Contract No. DE-AC02-05CH1123, by the National Energy Research Scientific Computing Center, a DOE Office of Science User Facility under the same contract; and by the U.S. National Science Foundation, Division of Astronomical Sciences under Contract No. AST-0950945 to NOAO.

The Pan-STARRS1 Surveys (PS1) and the PS1 public science archive have been made possible through contributions by the Institute for Astronomy, the University of Hawaii, the Pan-STARRS Project Office, the Max-Planck Society and its participating institutes, the Max Planck Institute for Astronomy, Heidelberg and the Max Planck Institute for Extraterrestrial Physics, Garching, The Johns Hopkins University, Durham University, the University of Edinburgh, the Queen's University Belfast, the Harvard-Smithsonian Center for Astrophysics, the Las Cumbres Observatory Global Telescope Network Incorporated, the National Central University of Taiwan, the Space Telescope Science Institute, the National Aeronautics and Space Administration under Grant No. NNX08AR22G issued through the Planetary Science Division of the NASA Science Mission Directorate, the National Science Foundation Grant No. AST–1238877, the University of Maryland, Eotvos Lorand University (ELTE), the Los Alamos National Laboratory, and the Gordon and Betty Moore Foundation.

This research has made use of data from the MOJAVE database that is maintained by the MOJAVE team \citep{Lister2018}. This research has made use of the SIMBAD database, operated at CDS, Strasbourg, France \citep{Wenger2000}.

We use the following software and packages throughout the work: APLpy \citep{R&B2012}, AstroPy \citep{Astropy12, Astropy18, Astropy22}, Astroquery \citep{G&S2019}, Matplotlib \citep{Hunter2007}, NumPy \citep{Walt2011, Harris2020}, Pandas \citep{McKinney2010, Reback2022}, SAOImage DS9 \citep{Joye2003}, SciPy \citep{Virtanen2022} and TOPCAT \citep{Taylor2005}. ChatGPT, a variant language model based on GPT-3 \citep{Brown2020} is employed to correct grammatical errors and impove readability within the text of this work.

\appendix
\label{append}

Here, we show the VLASS2 images of all the Roma-BzCAT sources that our algorithm classified as \textit{2-SIDE SEPARATED} or \textit{2-SIDE EXTENDED} (i.e., likely incompatible with being blazars).  
There are 36 sources classified as \textit{2-SIDE SEPARATED} and 105 sources classified as \textit{2-SIDE EXTENDED}; each source was visually inspected. For every source where the visual inspection led to a classification change, we assigned a visual flag value of 1 and provided our revised visual classification in the on-line available resulting table. In Fig.~\ref{fig:2ss_1}-\ref{fig:2se_4}, the sources with revised classifications (`VClass', refer to the result table) are marked with a star in their corresponding image. 

In Fig.~\ref{fig:mojave}, we show the 22 VLASS2 images of MOJAVE sources (see Section~\ref{sec:mojave}) classified with non-blazar-like morphology. However, 11 of them are consistent with COMPACT via visual inspection (marked with stars in Fig.~\ref{fig:mojave}). 
Nevertheless, all of the remaining 11 sources with non-blazar-like morphology are likely real blazars given that all of them have reported apparent superluminal motions in the MOJAVE database \citep{Lister2019}, suggesting a jet seen at small inclination to the line of sight. 
This demonstrates that even in the case of extended morphology, we cannot rule out 100\% the blazar nature of the sources, and other characteristics must also need to be taken into account.

\begin{figure*}
  \centering
  \makebox[\textwidth][c]{\includegraphics[scale=0.6]{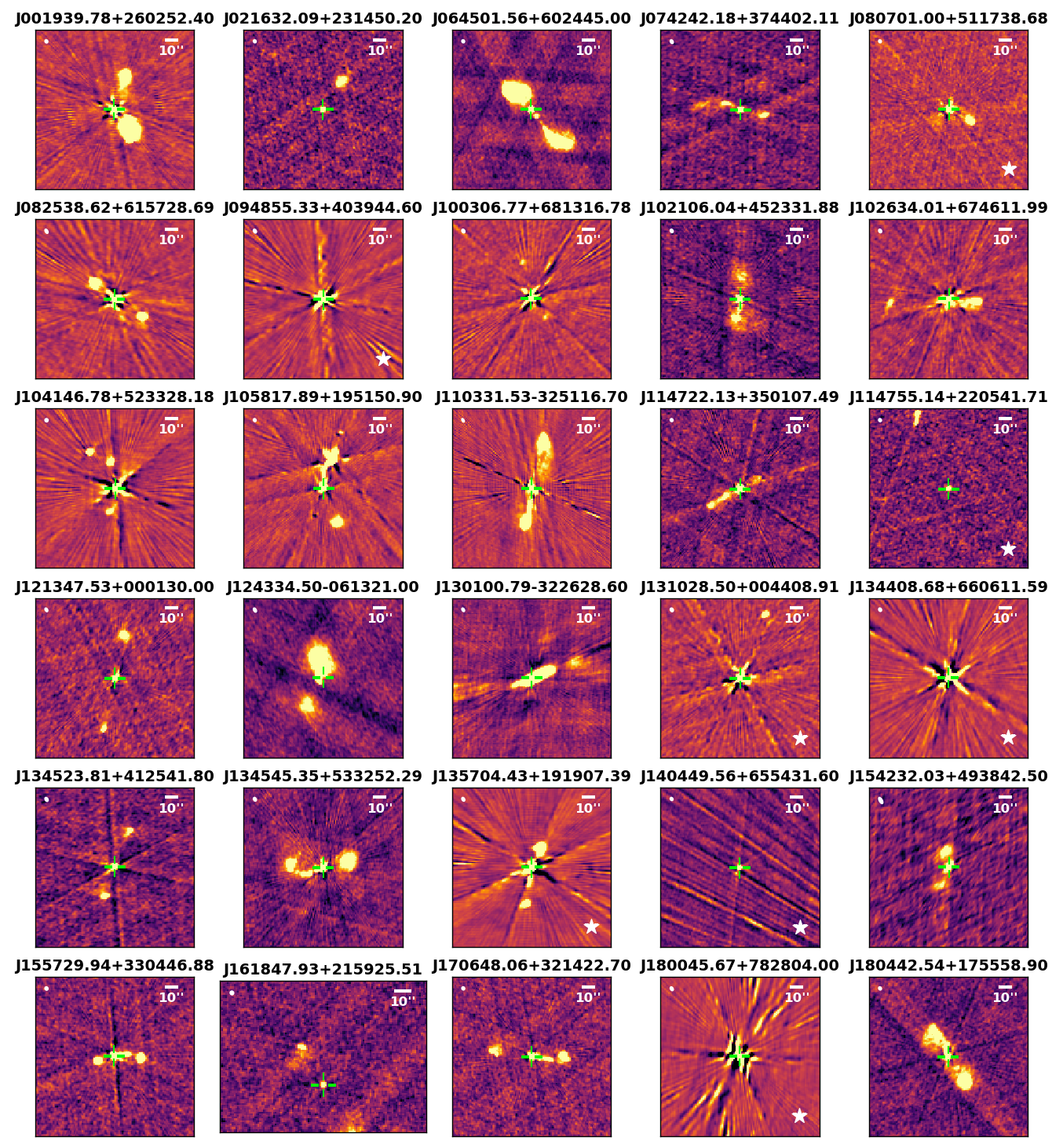}}
  \caption{All \textit{2-SIDE SEPARATED} sources with their VLASS2 image in Roma-BzCAT. Each source has undergone visual inspection, and any discrepancies found during this process are indicated by a visual flag in the resulting table. For each of these sources, our revised visual classification is provided. These revised sources are marked with a white star on their image.}
  \label{fig:2ss_1}
\end{figure*}

\begin{figure*}
  \centering
  \makebox[\textwidth][c]{\includegraphics[scale=0.6]{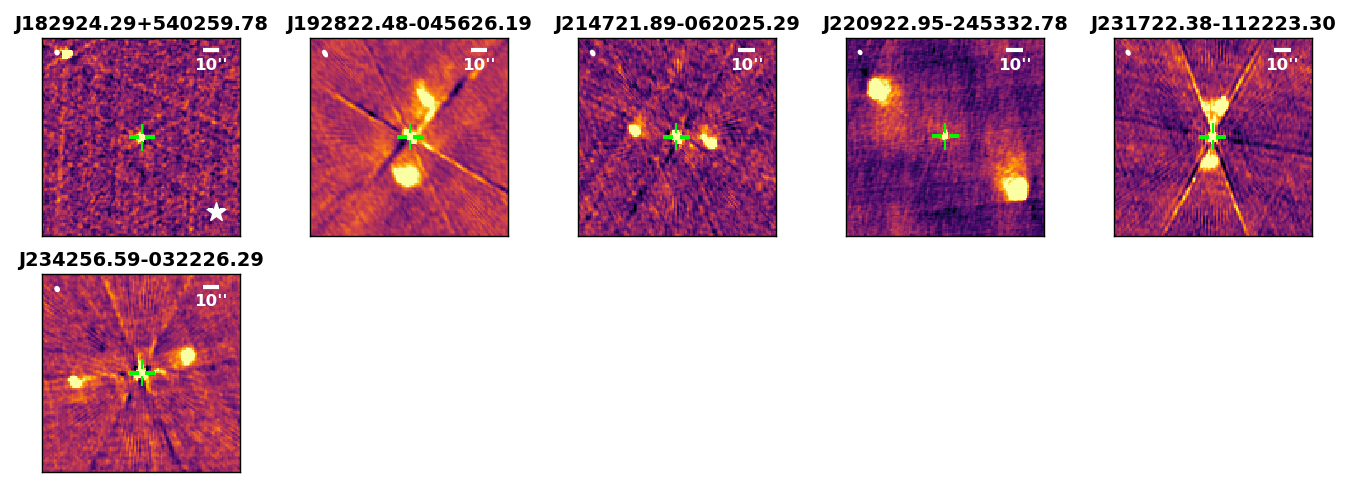}}
  \caption{Continued for Fig.~\ref{fig:2ss_1}.}
  \label{fig:2ss_2}
\end{figure*}

\begin{figure*}
  \centering
  \makebox[\textwidth][c]{\includegraphics[scale=0.6]{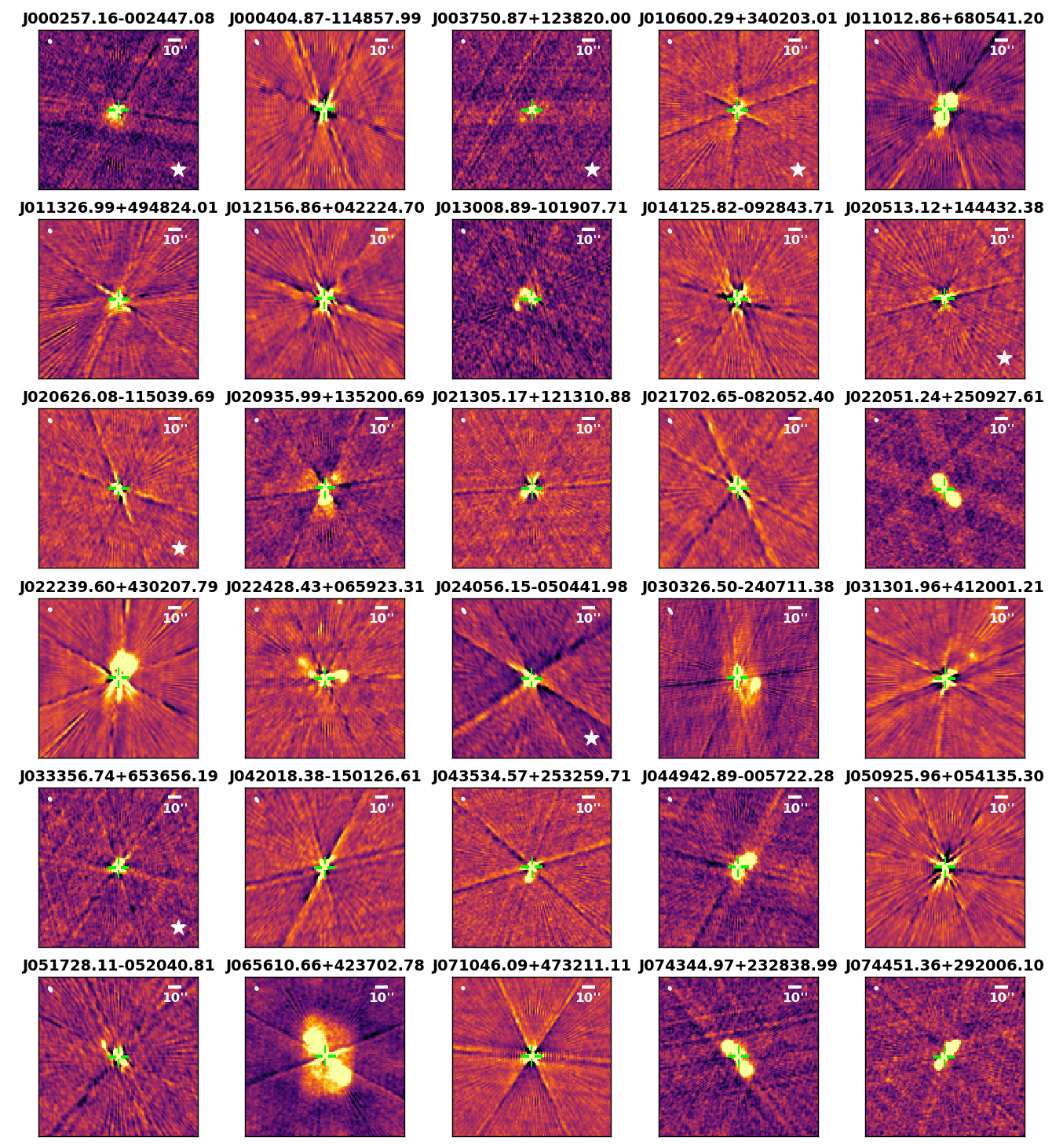}}
  \caption{Same as Fig.~\ref{fig:2ss_1} for \textit{2-SIDE EXTENDED} sources in Roma-BzCAT.}
  \label{fig:2se_1}
\end{figure*}

\begin{figure*}
  \centering
  \makebox[\textwidth][c]{\includegraphics[scale=0.6]{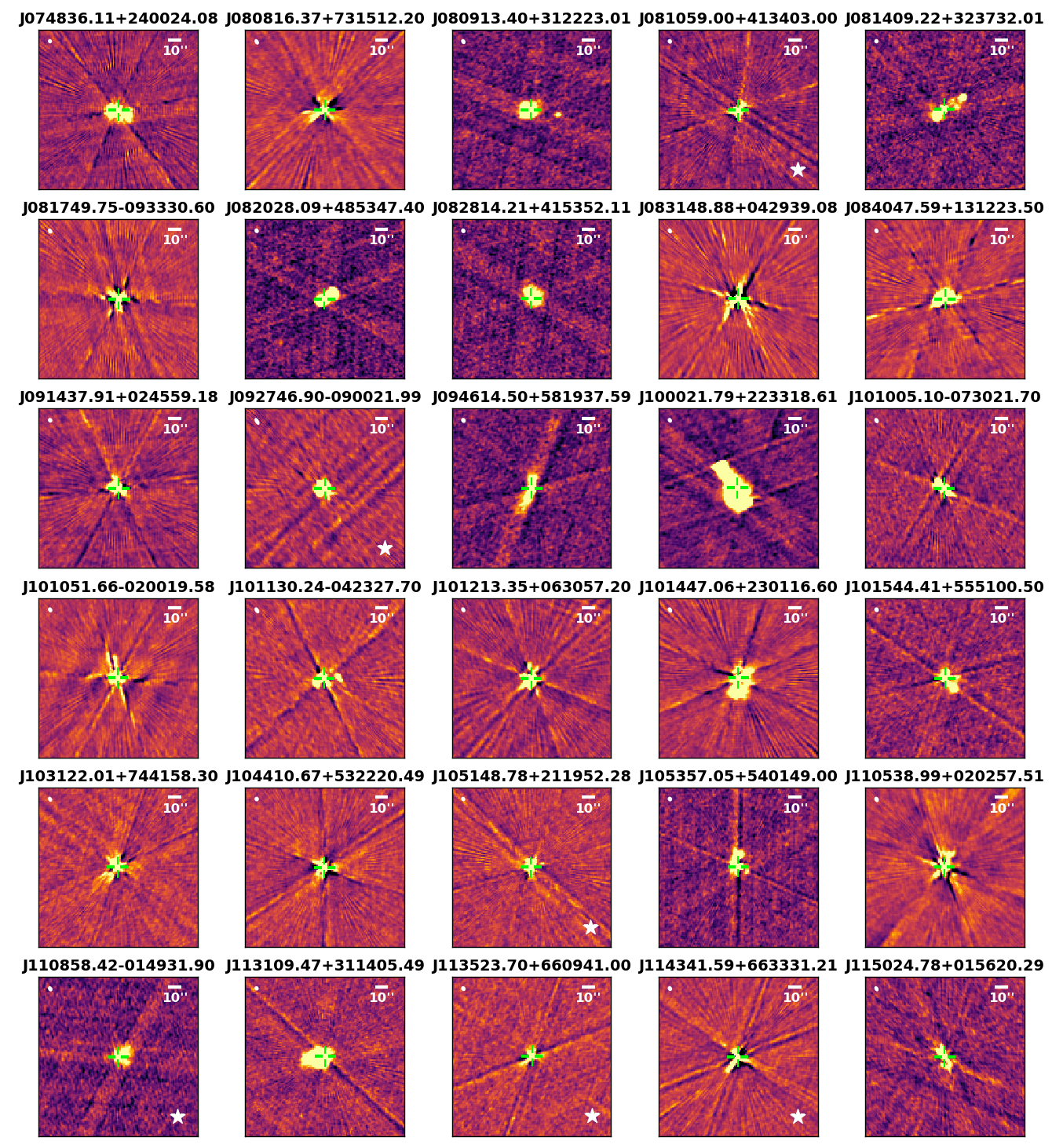}}
  \caption{Continued for Fig.~\ref{fig:2se_1}.}
  \label{fig:2se_2}
\end{figure*}

\begin{figure*}
  \centering
  \makebox[\textwidth][c]{\includegraphics[scale=0.6]{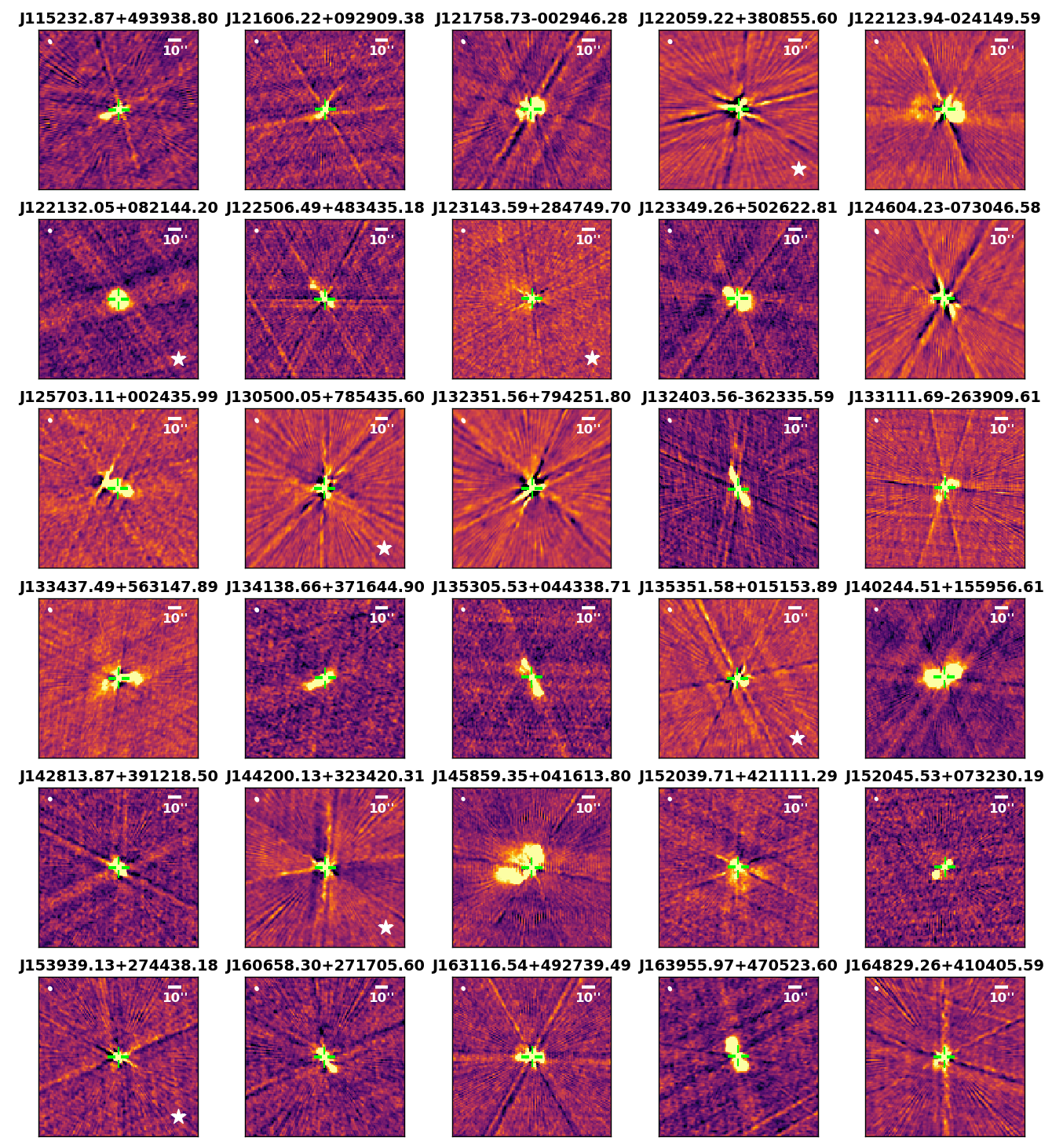}}
  \caption{Continued for Fig.~\ref{fig:2se_2}.}
  \label{fig:2se_3}
\end{figure*}

\begin{figure*}
  \centering
  \makebox[\textwidth][c]{\includegraphics[scale=0.6]{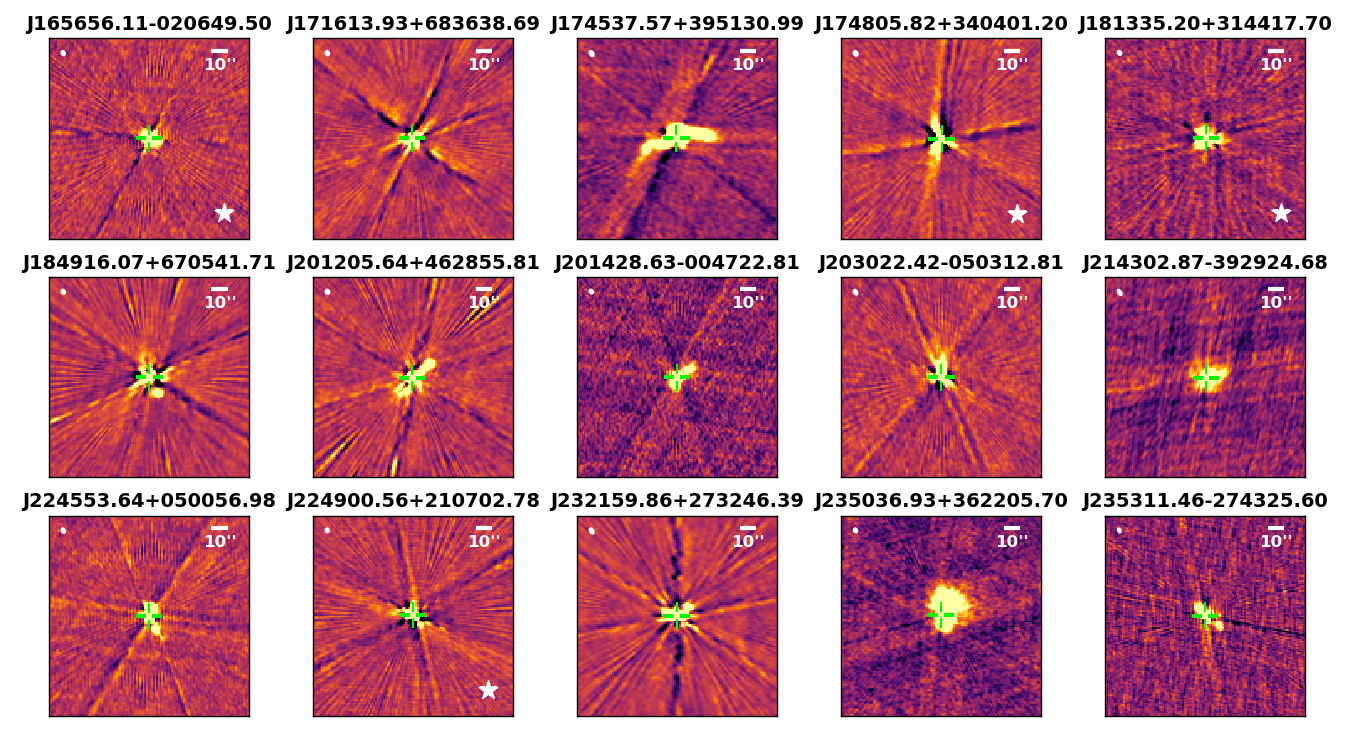}}
  \caption{Continued for Fig.~\ref{fig:2se_3}.}
  \label{fig:2se_4}
\end{figure*}

\begin{figure*}
  \centering
  \makebox[\textwidth][c]{\includegraphics[scale=0.6]{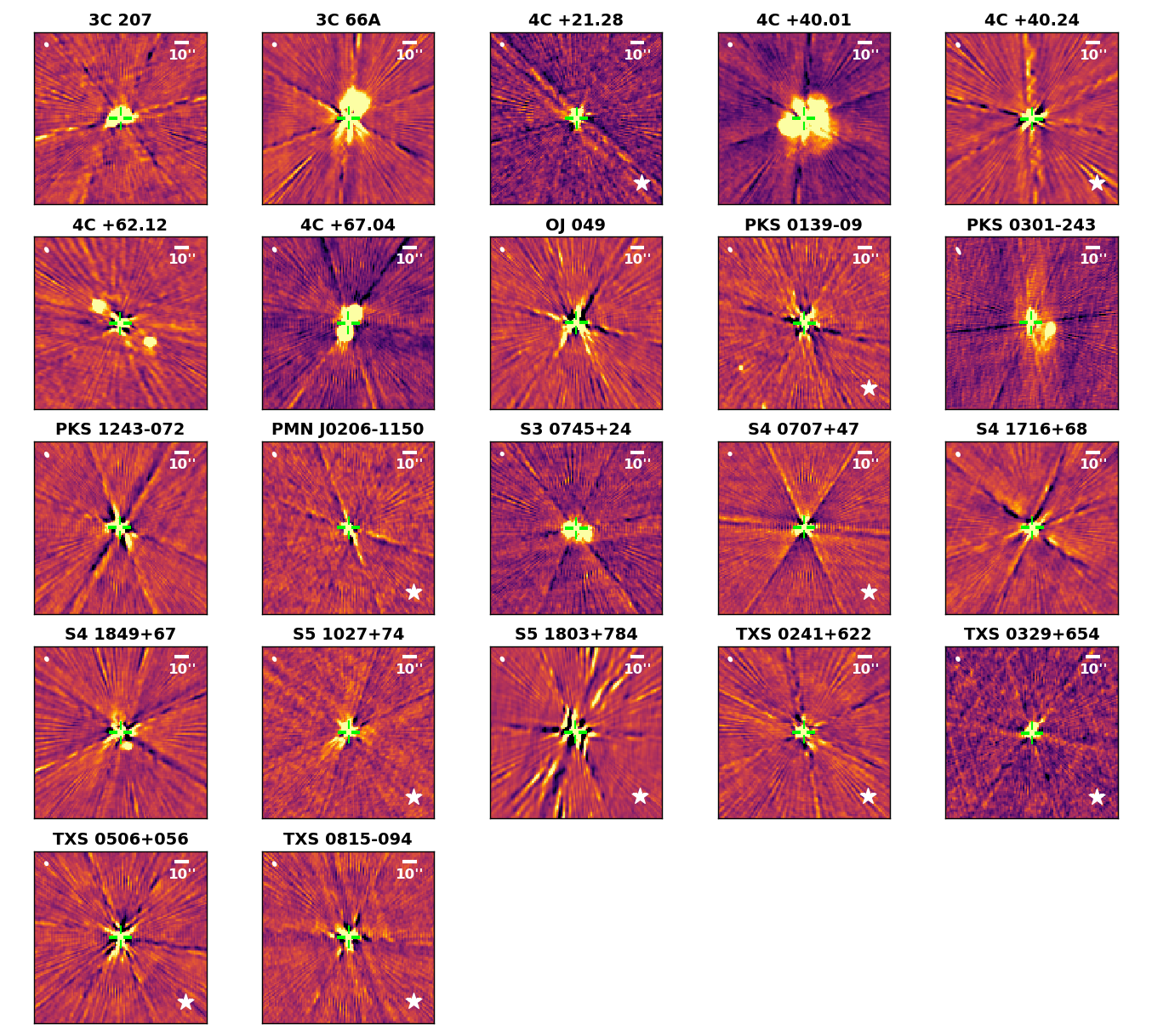}}
  \caption{VLASS2 images for 22 MOJAVE blazars classified as likely non-blazars. Each source has undergone visual inspection, and any discrepancies (are \textit{COMPACT} instead of non-\textit{COMPACT}) found during this process are indicated by a visual flag in the resulting table. For each of these sources, our revised visual classification is provided. These revised sources are marked with a white star on their image.}
  \label{fig:mojave}
\end{figure*}

\bibliographystyle{aasjournal}
\bibliography{sample631}{}

\end{CJK*}
\end{document}